\documentclass[conference]{IEEEtran}
\IEEEoverridecommandlockouts

\usepackage{cite}
\usepackage{amsmath,amssymb,amsfonts}
\usepackage{algorithmic}
\usepackage{graphicx}
\usepackage{textcomp}
\usepackage{xcolor}
\usepackage[hyphens]{url}

\usepackage{xspace}
\usepackage{subcaption}
\usepackage{fancyhdr}
\usepackage[bookmarks=true,breaklinks=true,letterpaper=true,colorlinks,citecolor=blue,linkcolor=blue,urlcolor=blue]{hyperref}

\def\BibTeX{{\rm B\kern-.05em{\sc i\kern-.025em b}\kern-.08em
    T\kern-.1667em\lower.7ex\hbox{E}\kern-.125emX}}

\newcommand{\name}{KHEPRI\xspace}
\newcommand{\speedupAppsInComputeCores}{43\%\xspace}
\newcommand{\speedupAppsInMemoryCores}{25\%\xspace}
\newcommand{\resultPerformance}{9.2\%\xspace}
\newcommand{\resultFPS}{7.3\%\xspace}
\newcommand{\resultEnergy}{4.8\%\xspace}
\newcommand{\resultPerformanceVertexStage}{0.8\%\xspace}

\begin{document}

\pdfpagewidth=8.5in
\pdfpageheight=11in


\pagenumbering{arabic}

\title{Design of a GPU with Heterogeneous Cores for Graphics}
\author{\IEEEauthorblockN{Aurora Tom\'{a}s\IEEEauthorrefmark{1}, Juan L. Arag\'{o}n\IEEEauthorrefmark{2}, Joan-Manuel Parcerisa\IEEEauthorrefmark{1}, Antonio Gonz\'{a}lez\IEEEauthorrefmark{1}}
\IEEEauthorblockA{
\IEEEauthorrefmark{1} \textit{Universitat Polit\`{e}cnica de Catalunya}
\\
}
\IEEEauthorblockA{
\IEEEauthorrefmark{2} \textit{Universidad de Murcia}
\\
}
}

\maketitle
\thispagestyle{plain}
\pagestyle{plain}


\begin{abstract}

Heterogeneous architectures can deliver higher performance and energy efficiency than symmetric counterparts by using multiple architectures tuned to different types of workloads. 
While previous works focused on CPUs, this work extends the concept of heterogeneity to GPUs by proposing \name, a heterogeneous GPU architecture for graphics applications.

Scenes in graphics applications showcase diversity, as they consist of many objects with varying levels of complexity. As a result, computational intensity and memory bandwidth requirements differ significantly across different regions of each scene. To address this variability, our proposal includes two types of cores: cores optimized for high ILP (compute-specialized) and cores that tolerate a higher number of simultaneously outstanding cache misses (memory-specialized). A key component of the proposed architecture is a novel work scheduler that dynamically assigns each part of a frame (i.e., a tile) to the most suitable core. Designing this scheduler is particularly challenging, as it must preserve data locality; otherwise, the benefits of heterogeneity may be offset by the penalty of additional cache misses. Additionally, the scheduler requires knowledge of each tile’s characteristics before rendering it. For this purpose, \name leverages frame-to-frame coherence to predict the behavior of each tile based on that of the corresponding tile in the previous frame.

Evaluations across a wide range of commercial animated graphics applications show that, compared to a traditional homogeneous GPU, \name achieves an average performance improvement of \resultPerformance, a throughput increase (frames per second) of \resultFPS, and a total GPU energy reduction of \resultEnergy. Importantly, these benefits are achieved without any hardware overhead.

\end{abstract}

\section{Introduction}
\label{sec:introduction}

Driven by the continuous demand for higher performance and the growing challenges of power consumption and heat dissipation, there has been increasing interest in recent years—both in industry and academia—in exploring various forms of architectural heterogeneity. Such interest arises from the potential to exploit execution diversity, allowing applications to identify and utilize processing cores best suited to their specific computational characteristics, thereby maximizing both performance and energy efficiency. Prior studies have investigated the use of application-specific accelerators \cite{intel2008sandy, amd2008future, nvidia2010benefits, nvidia2011smp}, as well as general-purpose cores featuring diverse microarchitectures \cite{kumar2003single, kumar2006core, kumar2004single, kumar2005heterogeneous, greenhalgh2011big, hill2008amdahl}, and heterogeneous systems supporting execution across different Instruction Set Architectures (ISAs) \cite{venkat2014harnessing, devuyst2012execution, venkat2016hipstr, venkat2019composite}. Collectively, these works demonstrate that, for many applications, architectural diversity provides superior adaptability and performance efficiency compared to homogeneous designs. However, while prior research has extensively explored heterogeneity in CPU-based systems, its implications and opportunities within GPU design remain largely unexplored.

Graphics applications usually display significant spatial heterogeneity because different regions of a frame present varying levels of detail, resulting in uneven computational workloads. For instance, some parts of a scene may consist primarily of background elements (e.g., the sky), whereas others may include numerous overlapping and highly detailed objects that demand substantially more work.

The continuous demand for visual realism has driven advancements in rendering quality and display refresh rates. Contemporary gaming applications employ increasingly complex geometry, characterized by a higher number of triangles per object, which substantially elevates the computational burden on GPU cores—commonly referred to as \textit{shader cores}. These cores execute user-defined programs to compute pixel colors through advanced lighting, shadowing, and texturing models. This stage of the graphics pipeline often constitutes a major performance bottleneck due to the growing complexity of shader programs and the significant memory bandwidth required for texture accesses.

To reduce memory pressure, Tile-Based Rendering (TBR) GPUs are the predominant architecture in mobile GPUs \cite{peddie2023history}. TBR efficiently exploits memory locality by dividing the screen into small regions, called \textit{tiles}, and processing them sequentially. This approach minimizes DRAM accesses by utilizing tile-sized on-chip buffers to store intermediate results. Given the inherent heterogeneity of frames, the resource demands of individual tiles can vary significantly. For instance, some tiles may require the application of richer textures, leading to an increased number of memory accesses and benefiting from higher Memory-Level Parallelism (MLP). In contrast, other tiles may rely more on computational power to apply lighting and shadowing models, for which exploiting larger Instruction-Level Parallelism (ILP) can be advantageous. One of the key motivations of this work is to address the varying resource requirements of different tiles during their execution.

In this paper, we harness the diversity offered by the divergence in memory accesses and computational demands among tiles within a frame. To address this variability, we leverage two types of GPU cores: cores optimized for handling memory-intensive workloads and cores designed for executing compute-intensive tasks. First, using a set of commercial graphics applications (see Section~\ref{sec:evaluation}), we analyze the most memory- and compute-intensive applications and, based on their requirements and characteristics, we define the microarchitecture for each GPU core type (see details in Section~\ref{sec:hetero}). We find that processing memory-intensive applications on a system configured with memory-specialized cores yields an average speedup of \speedupAppsInMemoryCores compared to a system with compute-specialized cores. Similarly, compute-intensive applications benefit from an average speedup of \speedupAppsInComputeCores when executed on a compute-specialized core system, as opposed to a memory-core system.

Similar to CPUs, designing a heterogeneous architecture is not trivial, as it requires effectively assigning workloads (in our case, the multiple tiles within a frame) to the most suitable hardware. This work proposes \name, a heterogeneous GPU architecture for rendering graphics applications that assigns compute-intensive tiles to compute-specialized cores and memory-intensive tiles to memory-specialized cores. In TBR GPU architectures, tiles can be processed in any order. Scheduling tiles to cores presents an important challenge since nearby tiles tend to share more textures than distant ones, owing to the natural temporal and spatial locality of textures. As a result, tile proximity becomes a critical factor to not penalize the effectiveness of cache memories. In consequence, \name features a tile scheduler that considers both tile proximity and workload affinity to assign tiles to cores.

To identify the core-type affinity of each tile, we need a method to predict the memory intensity that it will require before executing it. In animated applications, maintaining a high frame rate is essential for creating the illusion of movement. As a result, each frame is typically very similar to its predecessor, a well-known phenomenon referred to as \textit{frame-to-frame coherence} in the literature \cite{hubschman1982frame}. Our technique leverages this frame coherence to predict the memory intensity for each tile in a given frame based on statistics collected from the previous frame.

To summarize, in this paper we propose a novel GPU heterogeneous architecture for rendering animated graphics applications.
To the best of our knowledge, this is the first work to explore the use of heterogeneous cores in GPUs.
Our work makes the following key contributions:

\begin{itemize}
    \item We propose \name, a novel GPU architecture that leverages the inherent heterogeneity of graphical applications to drive future GPU designs.
    \item We propose two shader core microarchitectures: one optimized for handling many memory accesses (memory-specialized) and another for performing many arithmetic operations (compute-specialized).
    \item We propose a novel tile scheduler that is aware of both tile-core affinity and memory locality to not increase the miss ratio in the private L1 caches and the shared L2 cache.
    \item We show that \name offers important benefits in terms of performance (\resultPerformance improvement), frame rate (\resultFPS increase in frames per second), and energy (\resultEnergy reduction in total GPU energy consumption), without increasing the amount of hardware required.
\end{itemize}

The rest of the paper is organized as follows. Section~\ref{sec:background} provides some background on GPUs. Section~\ref{sec:hetero} presents \name. Section~\ref{sec:evaluation} describes the evaluation methodology. Section~\ref{sec:results} presents our experimental results and analysis. Section~\ref{sec:relatedwork} reviews some related work. Finally, Section~\ref{sec:conclusions} summarizes the main conclusions of this work.

\begin{figure*}[ht!]
    \centering
    \includegraphics[width=1\linewidth]{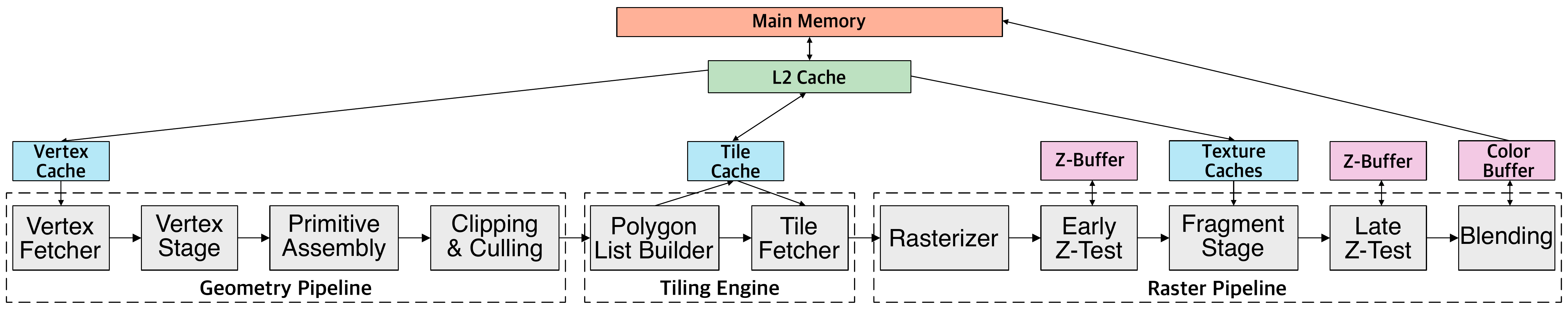}
    \caption{The Graphics Pipeline of a TBR GPU.}
    \label{fig:graphics-pipeline-gpu}
\end{figure*}

\section{Background}
\label{sec:background}

Contemporary mobile GPUs typically employ Tile-Based Rendering (TBR) architectures to minimize main memory accesses. This rendering paradigm, originally proposed to facilitate parallel rendering by ensuring non-overlapping tiles in the scene \cite{fuchs1989pixel, molnar1994sorting}, has become widely adopted in low-power and memory-bandwidth-constrained graphics systems. TBR divides the screen space into a grid of smaller rectangular regions of adjacent pixels, called \textit{tiles}, which are small enough to allow numerous operations on tile-sized on-chip buffers. Since off-chip memory is usually the primary source of power consumption in GPUs \cite{catthoor1994global, cozzi2012opengl, fromm1997energy}, these on-chip buffers significantly reduce power-hungry accesses to external memory, thereby saving both power and memory bandwidth. Antochi et al. \cite{antochi2004memory} show that TBR substantially reduces external data traffic compared to traditional, non-tile-based architectures, often referred to as Immediate-Mode Rendering (IMR) GPUs.

\subsection{Graphics Pipeline}

\autoref{fig:graphics-pipeline-gpu} illustrates the main stages of the Graphics Pipeline and the memory hierarchy organization in a TBR architecture. The rendering process typically involves two major pipelines: Geometry and Raster. The Geometry Pipeline performs geometry-related operations on the triangles that define the objects in the scene, generating the corresponding \textit{primitives} that lie within the visible frustum. The Raster Pipeline, in turn, discretizes each primitive into pixel-sized \textit{fragments}, which are then shaded and blended to produce their final output color for the screen image.

In TBR architectures, the Raster Pipeline renders \textit{tiles} instead of the entire frame to improve locality by keeping most data on-chip. To enable this tiling process, the geometry is first sorted into tiles, which are then processed by the rasterization stages. Classified as sort-middle in the literature \cite{real-time-rendering, molnar1994sorting}, these architectures rely on an intermediate phase called the \textit{Tiling Engine} to perform this sorting. As a result, TBR architectures feature three main pipelines, as shown in \autoref{fig:graphics-pipeline-gpu}.

The Geometry Pipeline is triggered by \textit{draw calls}, which are commands that request the rendering of a batch of objects. The Vertex Fetcher fetches the objects' vertices from memory, after which the Vertex Processors transform these vertices by executing a user-defined \textit{vertex shader} program. Once processed, the vertices are assembled in program order to form polygons, typically triangles. Next, for each primitive, it is determined whether it lies within the frustum, based on the camera's viewpoint. This Culling process discards primitives that are entirely outside the viewing volume. If a triangle is partially visible, a Clipping operation is applied, where the primitive is split into smaller triangles, and only those fully contained within the visible region are kept. The resulting primitives are then passed as input data to the Tiling Engine for further processing.

The Polygon List Builder is responsible for binning the primitives into tiles, i.e., generating a list in program order for each tile that includes all primitives fully (or partially) contained within it. These per-tile primitive lists are stored in a memory region called the \textit{Parameter Buffer}. Once all geometry has been processed and binned into tiles, the Tile Fetcher operates on a tile-by-tile basis, fetching the primitives associated with the current working tile. These primitives are then passed as input to the Raster Pipeline.

The Raster Pipeline renders tiles sequentially. The Rasterizer determines which pixels are covered by each primitive in the current tile and discretizes them into a set of \textit{fragments}. Additionally, the Rasterizer interpolates the attributes of the primitive. These fragments are grouped into 2x2 adjacent fragments to form \textit{quads}, which are sent to the Early Z-Test stage. The purpose of this stage is to eliminate fragments that are occluded by previously processed fragments. This is achieved by using a tile-sized on-chip buffer called the \textit{Z-Buffer}, which stores the depth value of the closest fragment processed so far for each tile's pixel position. The \textit{presumably} visible quads proceed to the Fragment Stage, where the shader cores compute the color for each fragment by executing user-defined \textit{fragment shader} programs, which apply a lighting model and its corresponding textures. Finally, the output colors are processed by the Blending Unit, which combines them with those already stored in the \textit{Color Buffer} at the same pixel position, considering transparency effects. In certain cases, the shader cores may modify the depth values of the fragments. In such situations, the Early Z-Test is bypassed, and visibility is determined post-shading (Late Z-Test).

Finally, once all primitives in the current tile have been fully rendered, the contents of the Color Buffer are flushed to the \textit{Frame Buffer}, a region in main memory that stores the data to be displayed on the screen. As a result, the Color Buffer is written to main memory once per tile. After all tiles of a frame have been processed, the frame is ready for display. Note that both the Z-Buffer and Color Buffer are tile-sized, allowing them to be stored in on-chip memory, which significantly reduces the need for accesses to off-chip DRAM.

\subsection{Shader Core Microarchitecture}

A GPU consists of multiple processing cores, commonly referred to as Streaming Multiprocessors (SMs), Compute Units (CU) or shader cores. Graphics applications are inherently parallel, as the computations for each pixel are independent of one another. To maximize throughput and hide memory latency, shader cores are designed to be highly multithreaded, managing a large number of threads concurrently. GPUs utilize a Single Instruction Multiple Threads (SIMT) architecture, which allows multiple threads to execute simultaneously on Single Instruction Multiple Data (SIMD) hardware \cite{lindholm2008nvidia}. SIMD lanes perform identical operations on different operands from distinct threads. The GPU hardware organizes these threads into groups called \textit{warps} \cite{lindholm2008nvidia} or \textit{wavefronts} in AMD GPUs, which are executed in lock-step. A warp represents the smallest unit of work, with all threads within a warp sharing the same instruction. The number of warps per shader core is implementation-specific.

\begin{figure}[b] 
    \centering
    \includegraphics[width=1\linewidth]{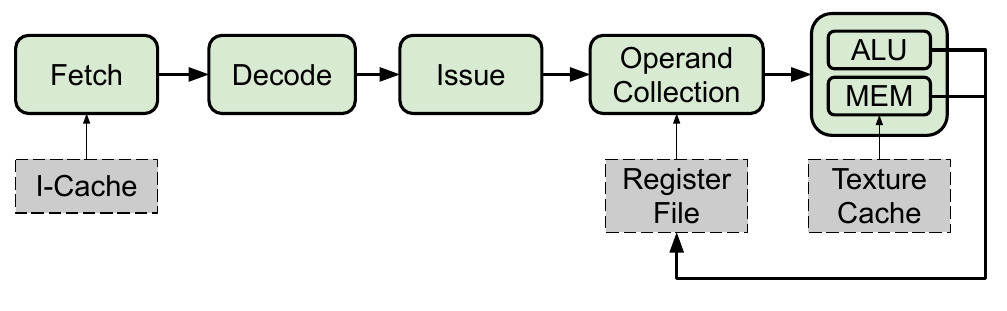}
    \caption{Shader core microarchitecture.}
    \label{fig:shader_core_diagram}
\end{figure}

Although vendors have not disclosed detailed information on the microarchitecture of shader cores in modern low-power TBR GPUs, we assume a plausible design based on prior academic research \cite{aamodt2018general, bakhoda2009analyzing, khairy2020accel, gpgpusim-manual} and publicly available diagrams \cite{frumusanu2019valhall}, as illustrated in \autoref{fig:shader_core_diagram}.

Each shader core can issue a specific number of instructions per cycle, depending on the architectural implementation. When a warp issues an instruction, it is first placed in a Collector Unit (CU), where it remains until all required source operands are retrieved from the Register File. To maximize Register File bandwidth, a shader core typically comprises multiple CUs. Once all source operands are available in the CU, the instruction is dispatched to the appropriate processing pipeline. The arithmetic pipeline includes several ALUs, which perform the required arithmetic operations. On the other hand, the memory pipeline (i.e., texture pipeline) handles address calculations, texture-related memory accesses, and filtering operations. Depending on the implementation, multiple memory accesses may be performed concurrently. After the instruction is processed, the result is written back to the Register File.

\subsection{Tile Scheduling}

The Tiling Engine determines the order in which tiles are processed. Since tiles operate independently, they can be executed in any order. However, primitives within each tile must be processed in program order to ensure correctness.

The two most common tile traversal orders in computer graphics are scanline order and Morton order \cite{morton1966computer}. The scanline order follows a row-major pattern, while Morton order follows a Z-shaped traversal. Although Morton order is more complex, it is more cache-friendly due to its ability to improve spatial locality. For this reason, we assume that the baseline GPU architecture in this work uses Morton order (also known as Z-order).

\subsection{Memory Organization}

\autoref{fig:graphics-pipeline-gpu} also depicts the memory hierarchy of a TBR GPU. Multiple L1 caches are used to store geometry (Vertex cache and Tile cache) and textures (Texture caches), all backed by a shared on-chip LLC, which is connected to the off-chip main memory. Additionally, there are local on-chip memories dedicated to storing the Z-Buffer and Color Buffer for the tile currently being processed. Notably, the Color Buffer directly transfers its contents to main memory once all the primitives of the current tile have been rendered. In contrast, the Z-Buffer does not need to be written back to main memory.

\section{KHEPRI: Heterogeneous GPU Architecture}
\label{sec:hetero}

In this paper, we propose \name, a novel GPU architecture that exploits intra-frame heterogeneity in graphics applications.
This is achieved by specializing the microarchitecture of the different shader cores to enhance the processing for specific workloads, such as those with higher computational demands or those that involve more memory accesses.
To determine whether a tile is executed on a compute-specialized or a memory-specialized shader core, \name incorporates a tile scheduler that effectively distributes tiles across the shader cores based on their characteristics while preserving spatial locality in the L1 and L2 caches and maintaining load balance across the shader cores.

\subsection{Specialized Shader Cores}

Graphics applications are naturally heterogeneous since computation intensity and memory demands vary significantly across regions within a single frame. To address this inherent diversity, we first define the microarchitectural parameters of the shader cores that are best suited for each type of task. 

\noindent \textbf{Compute cores.}
These cores are designed to handle compute-intensive tiles, whose performance depends mostly on the compute power rather than memory access efficiency. Shader programs executed on these tiles generally have a very high hit ratio in the L1 caches, suggesting that larger caches are unnecessary. Additionally, since memory instructions spend relatively little time in the memory pipeline, program dependencies are resolved sooner and can benefit from a higher instruction-level parallelism (ILP).
Thus, to satisfy the demands of these tiles, specialized compute cores feature more functional units (ALUs), increased issue width, and reduced cache overhead -- i.e., smaller L1 caches and a reduced number of Miss Status Holding Registers (MSHRs) -- compared to a modern mobile GPU core.

\noindent \textbf{Memory cores.}
In contrast, memory-intensive tiles spend a significant fraction of their execution time performing memory accesses. Their L1 cache miss ratio is much higher and, thus, memory accesses take much longer than in compute-intensive tiles. The primary goal when designing cores for memory-intensive tasks is to hide the long memory access latency. To achieve this, the number of warps per core is increased to allow a higher number of instructions on-the-fly to wait concurrently for the data being loaded from memory. To make this possible, a larger number of MSHRs is required to better exploit the underlying memory-level parallelism (MLP). In addition, larger cache sizes are needed to retain data locality. On the other hand, these cores do not require a high issue width since their ILP is limited.

The decision to assign each tile to the most appropriate core is crucial since a tile's performance can significantly improve when processed on a well-suited core, or degrade noticeably if mapped to the wrong one.
Our experiments showed that memory-intensive applications achieve an average speedup of \speedupAppsInMemoryCores when executed on memory-specialized cores, while compute-intensive applications experience an average speedup of \speedupAppsInComputeCores on compute-specialized cores.
That is, accurately identifying the adequate shader core type for a tile has a very important impact on overall performance, making this a key aspect of \name, which is discussed in detail next.

\subsection{Tile-Core Affinity}

In the Raster Pipeline, processing cores are organized into distinct Raster Units (RUs), with each RU responsible for processing a separate tile in parallel. This GPU architecture, which enables the concurrent processing of multiple tiles, is referred to as Parallel Tile Rendering (PTR). Tomás et al. \cite{tomas2024libra} have shown that PTR improves GPU resource utilization more effectively than dedicating all resources to processing a single tile. For this reason, we assume a PTR architecture as our baseline. 

Fragment shader programs are user-defined kernels executed on shader cores to compute the color of each fragment. These programs employ sophisticated lighting and shadowing models and apply textures (images mapped onto object surfaces to add high-frequency details). Due to the complexity of these programs and the memory demands of texture accesses, the Fragment Stage is typically the main bottleneck in the Graphics Pipeline.

\begin{figure}[t]
    \centering
    \includegraphics[width=1\linewidth]{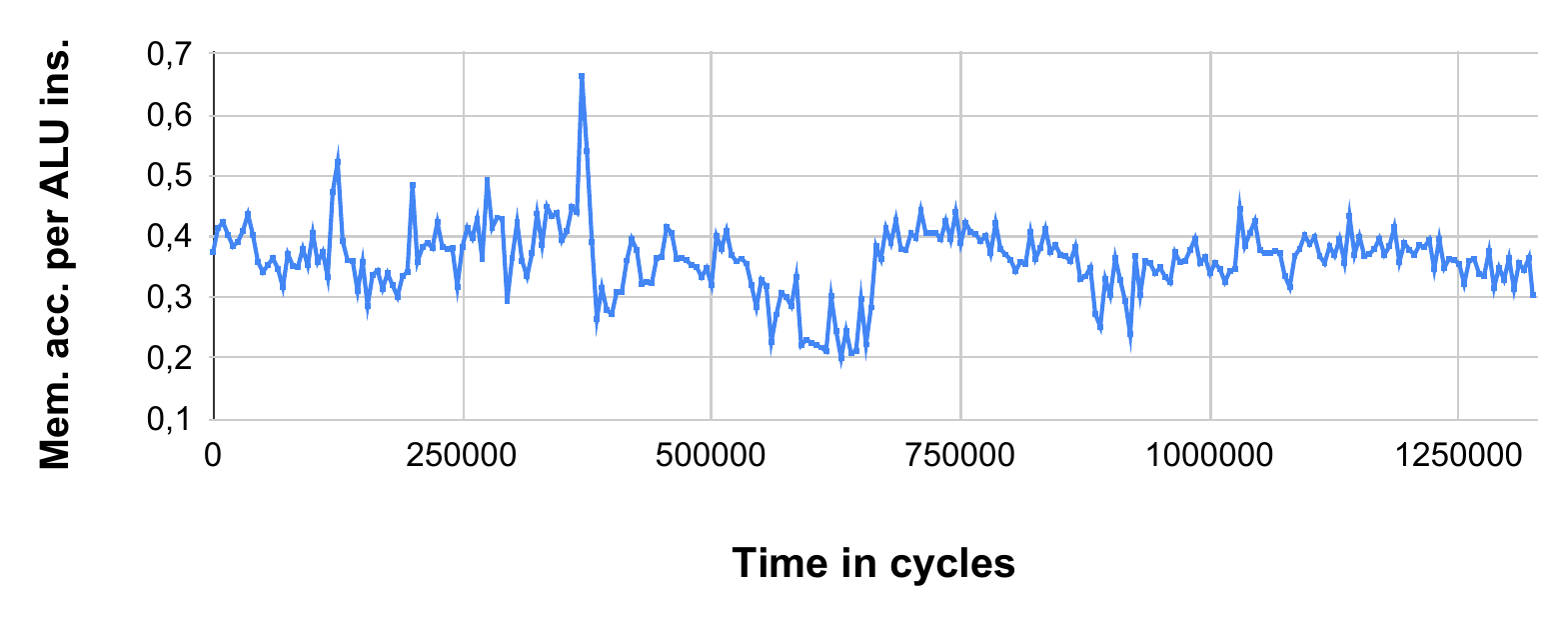}
    \caption{Ratio of memory requests to arithmetic instructions during the execution of a frame of Beach Buggy Racing (BBR).}
    \label{fig:timeline_bbr_alu_mem}
\end{figure}

\begin{figure}[t]
    \centering
    \includegraphics[width=1\linewidth]{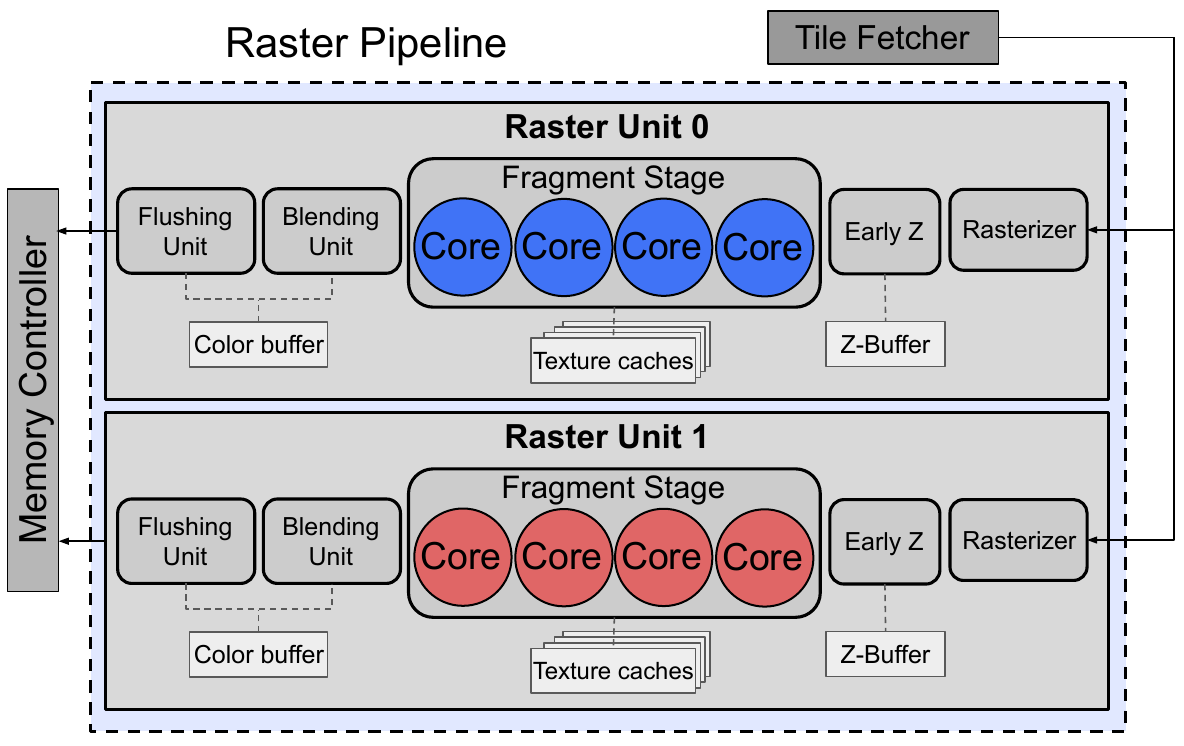}
    \caption{Architecture of \name. Each Raster Unit has its own private resources and specialized cores. Cores shaded in blue are compute-specialized, while cores shaded in red are memory-specialized.}
    \label{fig:ptr_organization}
\end{figure}

Since scenes are quite heterogeneous, different frame regions have varying resource requirements, particularly in terms of computation and memory bandwidth. 
As an example, \autoref{fig:timeline_bbr_alu_mem} shows the ratio of memory requests to arithmetic instructions during the rendering of a frame of the popular game Beach Buggy Racing (BBR). It can be observed that memory intensity varies significantly over time, with intervals of high memory-access demand interspersed with periods of lower memory activity.
To exploit this inherent heterogeneity, \name assigns tiles with higher memory requirements to memory-specialized cores, since they are capable of supporting a higher number of memory requests in parallel. In contrast, tiles with lower memory requirements are mapped to compute-specialized cores, as they make better use of their wider issue width and additional functional units. In this way, each task (in our case, the processing of a tile) is assigned to the core type that best suits its needs. In order to do that, \name features one Raster Unit composed of compute-specialized cores and another Raster Unit composed of memory-specialized cores, as illustrated in \autoref{fig:ptr_organization}.

Due to the continuous motion of objects in a scene, graphics applications exhibit a high degree of frame-to-frame coherence, where consecutive frames are visually consistent. Consequently, when two frames are visually analogous, their per-tile processing requirements should also be nearly identical.
\name exploits this natural coherence to predict the typology of the tiles in the next frame to be rendered.
The decision to schedule a given tile on the compute- or the memory-specialized RU is based on the memory intensity exhibited by the tile in the previous frame.
We define memory intensity as the number of misses per 1000 instructions (MPKI), which serves as a proxy for assessing the memory demands of a tile.
Tiles are arranged from highest to lowest memory intensity to determine each tile's relative memory demands within a frame.
To ensure the scheduling is time-balanced across the two Raster Units, we use a partitioning heuristic based on execution time.
This method distributes tiles by starting from both ends of the ranking list, utilizing each tile’s Fragment Stage execution time (in cycles) from the previous frame to ensure that the total processing time on both RUs is approximately equal. Note that, since tiles are independent of each other, they can be processed in any order.
To make this possible, a small table sized according to the number of tiles in a frame is required (see the implementation details in Section \ref{subsec:hw_implementation}).

\subsection{Preserving Locality}
\label{subsec:pres_locality}

\begin{figure*}
\begin{subfigure}[b]{.33\linewidth}
\centering
\includegraphics[width=\linewidth]{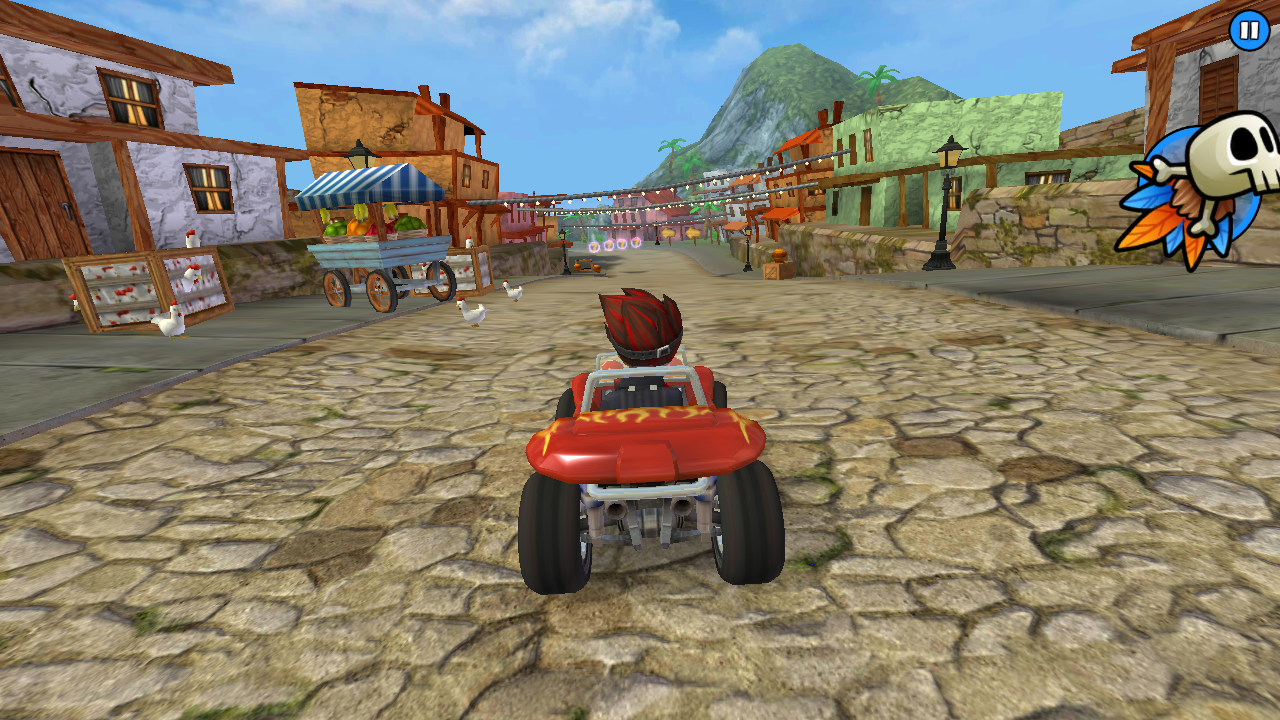}
\caption{Rendered frame.}
\end{subfigure}\hfill
\begin{subfigure}[b]{.33\linewidth}
\centering
\includegraphics[width=\linewidth]{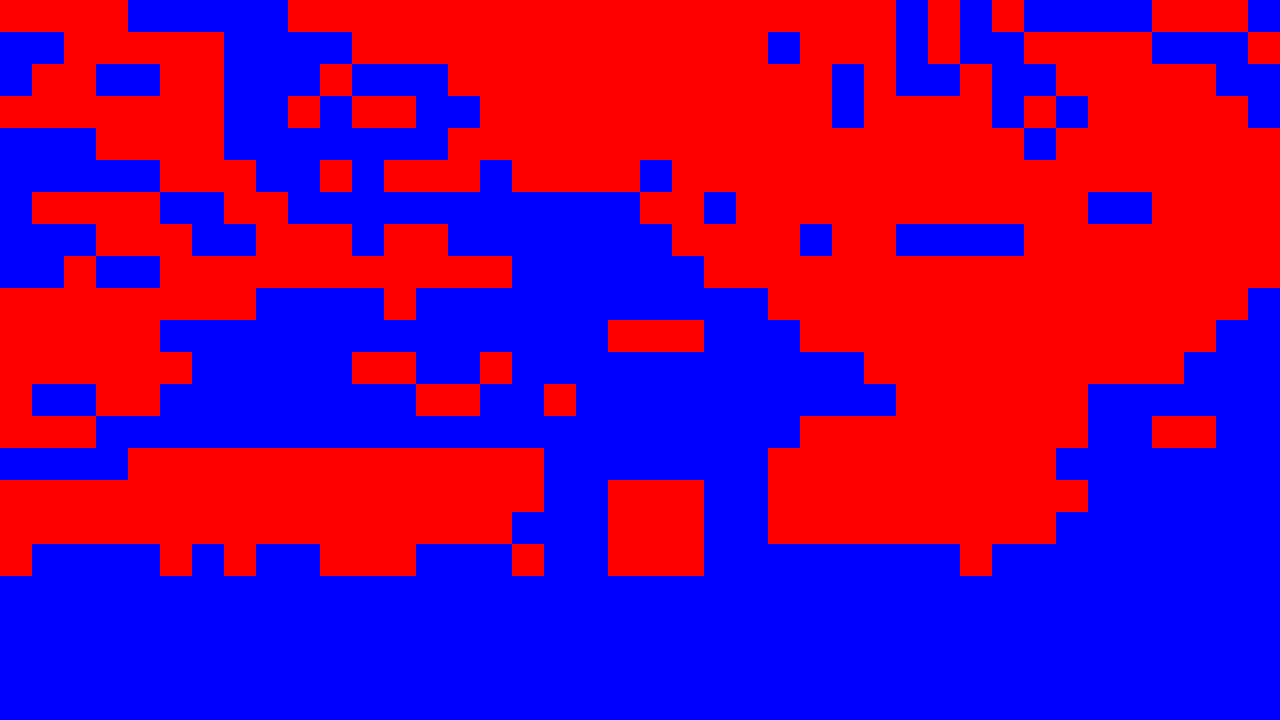}
\caption{Memory-intensity-based tile assignment.}
\end{subfigure}\hfill
\begin{subfigure}[b]{.33\linewidth}
\centering
\includegraphics[width=\linewidth]{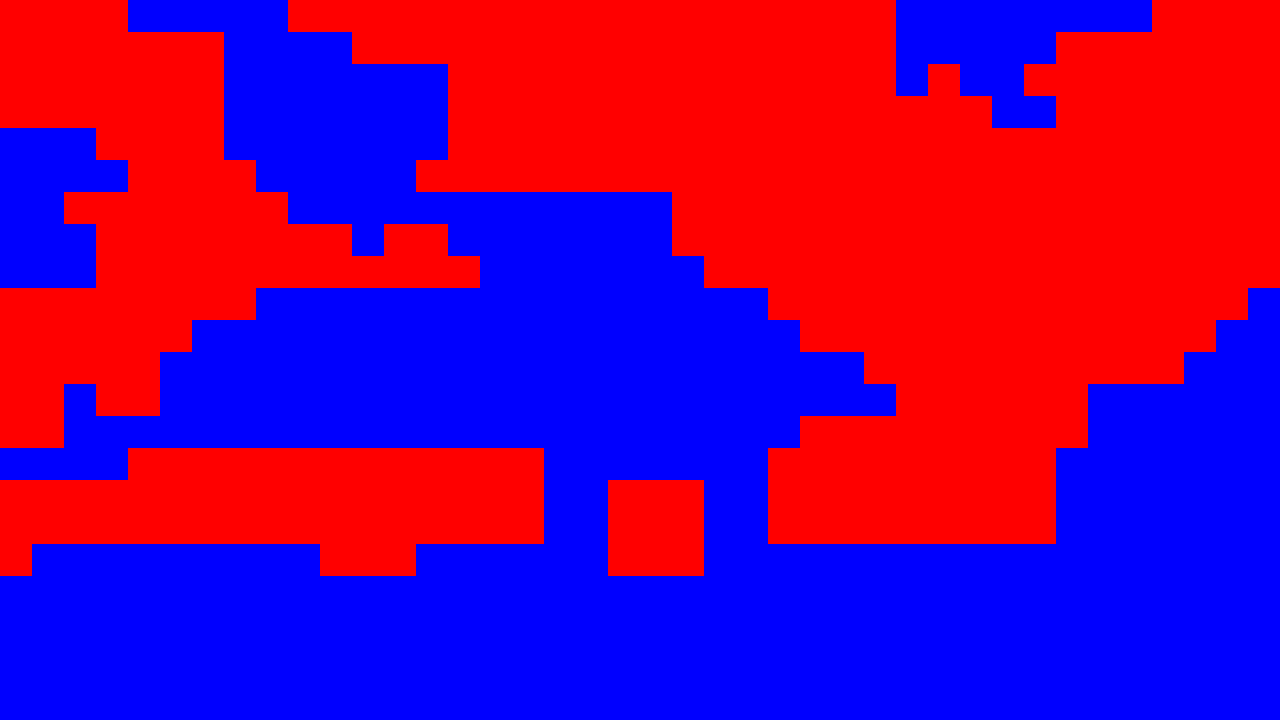}
\caption{Affinity- and locality-aware tile assignment.}
\end{subfigure}%
\caption{Beach Buggy Racing (BBR) \cite{game-bbr} rendered frame with corresponding core type assignment maps: (b) memory-intensity-only tile assignment, (c) affinity- and locality-aware tile assignment. Tiles shaded in blue are assigned to the compute-specialized cores, while tiles shaded in red are assigned to the memory-specialized cores.}
\label{fig:example-bbr-regions}
\end{figure*}

Data fetching from main memory is one of the major sources of energy inefficiency and often leads to performance bottlenecks in the Graphics Pipeline. In mobile GPUs, texture accesses dominate main memory traffic, making texture data locality essential for GPU efficiency, provided there are no bottlenecks in other pipeline stages. Notably, textures exhibit high locality at tile borders because adjacent tiles with shared edges are likely to access the same texture blocks in memory. That is, nearby tiles tend to reuse texture memory blocks. Similarly, spatially closer tiles also exhibit higher locality in the Parameter Buffer, as primitives are likely to overlap neighbor tiles, further emphasizing the importance of locality.

Assigning tiles to the RUs based solely on memory intensity could lead to the processing of scattered areas within a frame and the loss of locality. To illustrate this effect, \autoref{fig:example-bbr-regions}.b shows a tile assignment determined only by memory intensity for the popular game Beach Buggy Racing (BBR), revealing many isolated areas consisting of just one or a few tiles. These isolated regions of very few tiles imply that the corresponding RU would ``jump" from a non-neighbor tile when this region is started, and would jump to a non-neighbor tile when it is finished. These jumps between non-neighbor tiles would result in a significant loss of locality, particularly in the L2 cache. 

To address this, we propose an affinity- and locality-aware tile scheduling algorithm that balances core-type specialization (affinity) with spatial clustering (locality).

Tiles are initially assigned to cores according to their affinity for compute- or memory-specialized execution. Then, tiles that are surrounded predominantly by tiles of the opposite core type are reclassified. The algorithm distinguishes two types of tile isolation: (1) those that are \textit{totally isolated} (i.e., all four adjacent neighbors --- top, bottom, left, and right ---  belong to the other category); and (2) those that are \textit{highly isolated} (i.e., at least 75\% of their neighboring tiles are of the other core type). These tiles are subsequently considered for reassignment to the opposite category---that is, to the core type of the majority of their neighbors---while ensuring that the total number of cycles assigned to both RUs remains balanced. Tile reclassification proceeds only while there are candidate tiles available for swapping between the two core types (memory and compute), so that the resulting execution time for both RUs remains balanced. 
As a result of the reassignment, \textit{highly-isolated} tiles can create new \textit{totally isolated} tiles. Therefore, the algorithm performs a final pass to reidentify and reassign only these \textit{totally isolated} tiles to the majority type. Again, this tile reclassification proceeds only when there are candidate tiles of both types, so that the total execution time of each category remains balanced.

Finally, tiles are grouped into \textit{regions} of adjacent tiles surrounded by tiles of the opposite type. 
Regions that are too small are not desirable since the performance penalty from locality loss outweighs the gains from core specialization. For this reason, if a region contains fewer tiles than a defined threshold (8 tiles), it is merged with its surrounding region. 

To identify the distinct regions within a frame, we employ the Flood Fill algorithm, which is a well-known method used to detect connected tiles of the same type (i.e., regions). We traverse the connected regions employing the Breadth-First Search (BFS) algorithm to identify all connected tiles of the same core type. \autoref{fig:example-bbr-regions}.c illustrates the resulting tile assignment for the BBR application after applying the affinity- and locality-aware tile assignment algorithm. Compared to memory-intensity-only scheduling, the resulting regions are smoother and tiny regions do not exist, demonstrating improved locality retention without sacrificing the benefits of core specialization.

\begin{table}[b]
  \centering
  \caption{GPU simulation parameters.}
  \label{table:gpu-parameters}
    \begin{tabular}{ll}
      \hline
      & \textbf{Global Parameters} \\
      \hline
      Tech Specs & 800 MHz, 1V, 22nm \\
      Screen Resolution & 1920x1080 (Full HD) \\
      Tile Size & 32x32 pixels \\
      Raster Units & 2 \\
      Cores per Raster Unit & 4 \\
      \hline
      & \textbf{Main Memory (LPDDR4)} \\
      \hline
      Tech Specs & 1.6 GHz, 1.2V \\
      Latency & 50-100 cycles \\
      Size & 8 GB \\
      \hline
      & \textbf{Caches} \\
      \hline
      Vertex Cache & 64-bytes/line, 4KB, 2-way, 1 cycle \\
      Tile Cache & 64-bytes/line, 32KB, 4-way, 2 cycles \\
      L2 Cache (shared) & 64-bytes/line, 2MB, 8-way, 18 cycles \\
      \hline
      & \hspace{5pt} \textbf{Baseline} \hspace{35pt} \textbf{\name}\\
      \hline
      Cores Configuration & \hspace{3pt} 8$\times$ Regular \hspace{20pt} 4$\times$ Compute + \\
       & \hspace{15pt} \hspace{50pt} 4$\times$ Memory \\
      \hline
    \end{tabular}
\end{table}

\subsection{Tile Scheduling}

The regions are processed in scanline order, as they appear in the frame. Memory-intensive regions are processed on memory-specialized cores, while compute-intensive regions are processed on compute-specialized cores. Since tiles are the unit of work for the RUs, \name's Tile Fetcher assigns an entire region to a single RU, scheduling its tiles one after another on that RU. This approach helps preserve texture locality within each RU.

Determining the tile traversal order within a region is non-trivial, as preserving spatial locality is desired to increase cache locality. We traverse tiles within a region following an S-order, i.e., row by row and alternating direction between consecutive rows. This approach ensures that consecutive tiles are almost always neighbors, while remaining simple and computationally efficient.

\section{Evaluation methodology}
\label{sec:evaluation}

\subsection{Simulation Infrastructure}

To evaluate our proposal, we employ TEAPOT \cite{arnau2013teapot}, a cycle-accurate GPU simulation framework that runs unmodified Android applications to assess the performance and energy consumption of the modeled GPU and memory system. In order to do that, TEAPOT relies on established tools such as McPAT \cite{li2009mcpat} for area and energy estimation, and DRAMsim3 \cite{li2020dramsim3} to model the timing of DRAM and memory controllers. 
\autoref{table:gpu-parameters} and \autoref{table:shader-core-parameters} show the GPU and shader core microarchitectural parameters used in our simulations, respectively. Note that the baseline homogeneous GPU core architecture is modeled to closely resemble a modern ARM Valhall mobile GPU \cite{arm-gpu-datasheet, arm-gpu-valhall}. Each shader core contains its own private texture cache.

\begin{table}[t]
  \centering
  \caption{Core type configurations.}
  \label{table:shader-core-parameters}
    \begin{tabular}{lccc}
      \hline
      & \textbf{Compute} & \textbf{Memory} & \textbf{Baseline} \\
      \hline
      \# Warps & 64 & 96 & 64 \\
      Issue width & 6 & 3 & 4 \\
      \# CUs & 18 & 12 & 12 \\
      \# ALUs & 5 & 3 & 4 \\
      \# Memory Pipelines & 2 & 2 & 2 \\
      Texture Cache size & 8KB & 32KB & 32KB \\
      \# MSHRs (per cache) & 32 & 128 & 128 \\
      \hline
    \end{tabular}
\end{table}

\begin{tiny}
\begin{table*}
  \centering
  \caption{Evaluated benchmarks.}
  \label{table:benchmarks}
  \fontsize{6}{6}\selectfont
    \begin{tabular}{lccccc|lccccc}
        \hline
        \textbf{Benchmark} & \textbf{Alias} & \textbf{Genre} & \textbf{Type} & \textbf{Downloads} & \textbf{Footprint} &
        \textbf{Benchmark} & \textbf{Alias} & \textbf{Genre} & \textbf{Type} & \textbf{Downloads} & \textbf{Footprint} \\
        & & & & \textbf{(millions)} & \textbf{(MB)} & & & & & \textbf{(millions)} & \textbf{(MB)} \\
        \hline
        Air Attack & AAt & Action & 2.5D & 10 & 2.1 & 
        Gravity Tetris & GrT & Puzzle & 3D & 5 & 0.8 \\
        
        Among Us & AmU & Action & 2.5D & 500 & 3.4 & 
        Gravity: Don't Let Go & Gra & Arcade & 3D & 1 & 5.2 \\
        
        Angry Birds & AnB & Puzzle & 2D & 100 & 1.1 &
        Hill Climb Racing & HCR & Racing & 2D & 1000 & 2.8 \\
        
        Archery Master 3D & Arc & Sports & 3D & 100 & 2.3 &
        Hot Wheels: Race Off & HoW & Racing & 2.5D & 50 & 8.3 \\
        
        Beach Buggy Racing & BBR & Racing & 3D & 10 & 7.5 &
        Jetpack Joyride & Jet & Arcade & 2D & 100 & 0.7 \\
        
        Block Blast! & BlB & Puzzle & 2D & 100 & 3.6 & 
        3D Maze / Labyrinth & Maz & Adventure & 3D & 10 & 3.6 \\
        
        Candy Crush Saga & CCS & Casual & 2D & 1000 & 2.6 &
        Plants vs. Zombies & PVZ & Strategy & 2D & 500 & 2.7 \\
        
        Captain America: Sentinel of Liberty & CAm & Action & 2.5D & 5 & 1.5 &
        Real Steel World Robot Boxing & RSt & Action & 3D & 50 & 4.9 \\
        
        City Racing 3D & CRa & Racing & 3D & 50 & 3.5 &
        Rise of Kingdoms: Lost Crusade & RoK & Strategy & 2.5D & 50 & 7.1 \\
        
        Clash of Clans & CoC & Strategy & 2.5D & 500 & 2.4 &
        Royal Match & RoM & Puzzle & 2.5D & 100 & 27.5 \\
        
        Counter Strike & CoS & Action & 3D & 50 & 0.7 &
        Sniper 3D: Gun Shooting Games & S3D & Action & 3D & 500 & 6.0 \\
        
        Crazy Snowboard & CrS & Sports & 3D & 15 & 0.9 &
        Sonic Dash & SoD & Arcade & 3D & 100 & 4.8 \\
        
        Derby Destruction Simulator & DDS & Racing & 3D & 10 & 3.2 &
        Subway Surfers & SuS & Arcade & 3D & 1000 & 2.9 \\
        
        Forest 2 & Fo2 & Adventure & 3D & 1 & 4.7 &
        Tetris & Tet & Puzzle & 2D & 10 & 9.4 \\

        Geometry Dash Lite & GDL & Rhythm & 2D & 100 & 1.2 &
        Tigerball & TiB & Casual & 3D & 10 & 6.5 \\

        Golf Battle & GoB & Sports & 3D & 50 & 4.9 &
        Vegas Crime Simulator & VCS & Action & 3D & 10 & 4.3 \\

        \hline
    \end{tabular}
\end{table*}
\end{tiny}

\subsection{Benchmarks}

To provide confident results for the evaluation of \name, we have selected a wide range of commercial Android graphics applications as benchmarks. The selection was based on two criteria: variety, to encompass a wide diversity of applications, and popularity, as measured by the number of downloads in the Google Play Store.

\autoref{table:benchmarks} shows the set of benchmarks used to evaluate our proposal.
Our evaluation covers games with diverse perspectives and dimensionalities, including 2D games (e.g., CCS), 2.5D games (e.g., CoC), and 3D games (e.g., SuS). Additionally, the different games show significant variability in their average memory footprint per frame. 
Across all benchmarks, the average memory footprint exceeds 4MB; however, applications such as HoW and RoM exhibit substantially higher memory requirements, whereas benchmarks like CrS and Jet have minimal memory demands.

\section{Experimental results}
\label{sec:results}

In this section, we evaluate and analyze \name in terms of performance, energy, and hardware overhead compared to a conventional GPU composed of multiple symmetric cores. Besides, we highlight the importance of employing an effective tile scheduler to fully exploit the benefits of a heterogeneous shader core architecture; otherwise, such a design could be counterproductive. Finally, we assess the implementation overheads.

\subsection{\name Results}
\label{subsec:khepri_results}

\begin{figure*}
    \centering
    \includegraphics[width=1\linewidth]{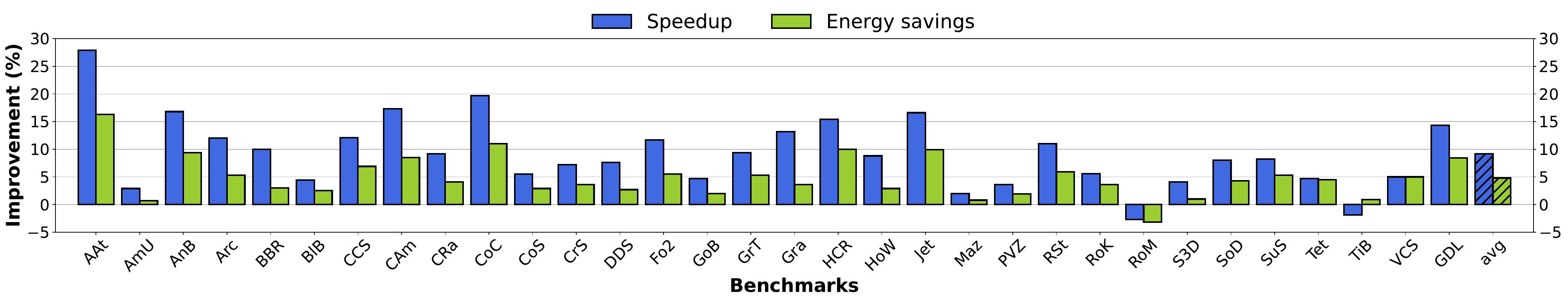}
    \caption{Speedup and energy savings of \name w.r.t. a traditional homogenous GPU.}
    \label{fig:khepri_speedup_energy}
\end{figure*}

\noindent \textbf{1) Performance.}
The blue bars in \autoref{fig:khepri_speedup_energy} show the speedup achieved by \name with respect to the baseline GPU, which consists of multiple homogeneous cores following a traditional microarchitecture, as detailed in \autoref{table:shader-core-parameters}. Note that the reported speedups correspond to the rasterization phase, which is where heterogeneity among tiles is significant and it is the most time-consuming part of the graphics pipeline. Overall, we observe an average speedup of \resultPerformance across the evaluated benchmarks. It can be seen that our approach is beneficial for all types of applications by reconsidering the architecture of GPUs and smartly scheduling the workload across the cores. For instance, AAt and CoC significantly improve performance by 28\% and 19.6\%, respectively. Only 2 out of the 32 benchmarks (RoM and TiB) showed a minimal slowdown, so we can conclude that \name generally provides good performance. Note that this is a meaningful improvement since performance is critical in real-time rendering, resulting in an average \resultFPS increase in frame rate (FPS), which corresponds to the execution of the entire graphics pipeline, including both the geometry and the rasterization stages. Performance improvements also yield higher-quality rendered outputs.

To better understand these results, let us highlight some examples. First, we discuss the applications least benefited by the proposal. TiB and AmU are two examples of applications in which most of their tiles are memory-intensive (e.g., they exhibit an average MPKI values in the L1 cache of 77 and 65, respectively). When most of the tiles are memory-intensive, some are necessarily mapped to the compute cores that offer a lower MLP, since this is better than leaving these cores unused. Our results show that these particular benchmarks are heavily penalized by the reduction in the global number of MSHRs, resulting in a nearly 20\% decrease in the number of concurrently handled misses. This inability to sustain additional memory accesses is particularly detrimental for memory-bound applications, limiting the capacity to hold requests on the fly. 

On the other hand, practically all tiles in RoM (13 MPKI) and Maz (15 MPKI) are compute-intensive. Once more, since it is better to utilize all shader cores, some tiles are necessarily assigned onto the memory-specialized cores, which are constrained by a lower instruction throughput, and cause a slight performance degradation.
Note, however, that these are extreme cases, as shown in \autoref{fig:khepri_speedup_energy}.

In general, we observe substantial benefits for the vast majority of applications that show a mix of both memory-intensive and compute-intensive tiles, proving the ability of \name to effectively leverage the varying resource demands of tiles and boost performance significantly. 

Additionally, note that the kernels corresponding to the geometry pipeline (vertex programs) are also executed on the shader cores. Since these programs focus on computation and modern GPUs employ a unified shader model architecture \cite{shirley2009fundamentals, real-time-rendering}, we schedule them on the compute-specialized cores. These applications are well-suited for such cores because they benefit from a wider issue width, and more functional units, while having small memory requirements. While the homogeneous architecture already provides good throughput for this part of the application, \name provides a minor additional speedup of \resultPerformanceVertexStage. Nevertheless, the geometry stages represent a very small fraction of the total execution time.


\begin{figure*}[ht!]
    \centering
    \includegraphics[width=1\linewidth]{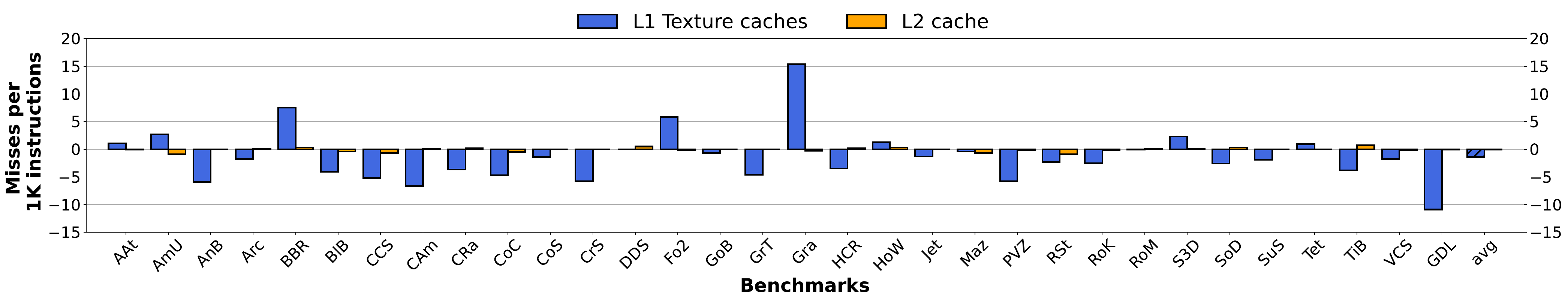}
    \caption{Number of misses per 1000 instructions in L1 texture caches and L2 cache w.r.t. a traditional homogenous GPU (the lower the better).}
    \label{fig:khepri_l1_l2_hit_ratio}
\end{figure*}

\vspace{3mm}
\noindent \textbf{2) Locality.}
\autoref{fig:khepri_l1_l2_hit_ratio} shows the number of misses per 1000 instructions for both the overall L1 texture caches (represented by the blue bars) and the L2 cache (represented by the orange bars), compared to the baseline.
It can be seen an average reduction of 1.4 misses per 1000 instructions compared to the baseline for L1 caches. In general, most applications remain unaffected, as expected, since the purpose of \name is to preserve locality. Some applications experience a reduction in the total miss count (e.g., GDL with 11 fewer misses per 1000 instructions), while others see an increase (e.g., Gra with 15 more misses per 1000 instruction).
Note, however, that a reduction in the miss count does not necessarily directly translate to improved performance, as bottlenecks may exist in other stages of the pipeline. For example, TiB improves locality by almost 4 misses per instruction, yet these gains do not result in performance improvements. As discussed in the previous analysis, this application is particularly constrained by a reduction in the overall number of MSHRs.

On the other hand, we observe that the locality for the L2 cache is hardly affected, with an average miss count decrease of only 0.1 misses per 1000 instructions. This is expected, as we do not explore techniques aimed at improving locality, which would be orthogonal to this work.


\begin{figure*}[ht!]
    \centering
    \includegraphics[width=1\linewidth]{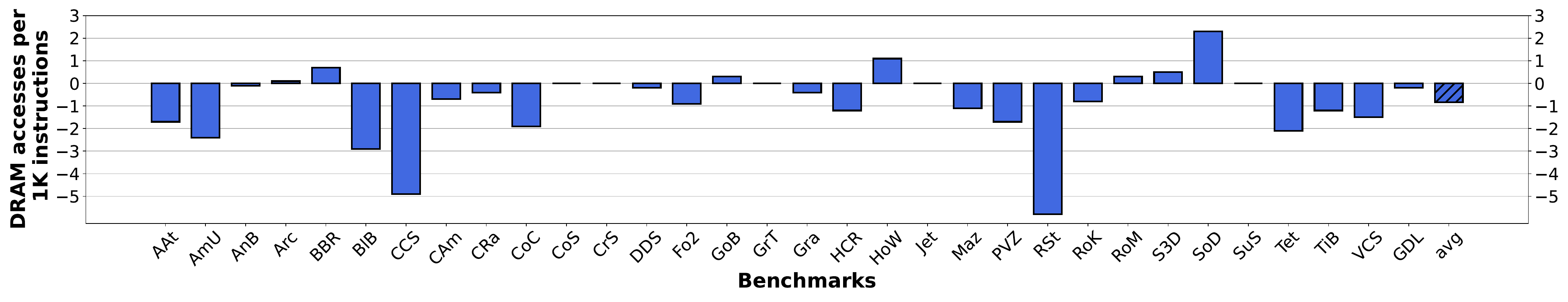}
    \caption{Number of main memory accesses per 1000 instructions w.r.t. a traditional homogenous GPU (the lower the better).}
    \label{fig:khepri_dram_accesses}
\end{figure*}

\vspace{3mm}
\noindent \textbf{3) Main Memory Accesses.}
\autoref{fig:khepri_dram_accesses} shows the number of main memory accesses per 1000 instructions compared to a baseline homogeneous GPU. On average, there is a decrease of 0.8 main memory accesses per 1000 instructions, which represents a 3.7\% reduction in read DRAM accesses, mainly due to the slight improvement in L1 texture cache hit rate.


\vspace{3mm}
\noindent \textbf{4) Energy.}
The green bars in \autoref{fig:khepri_speedup_energy} show the reduction in total energy with respect to the baseline GPU. Overall, \name achieves an average energy decrease of \resultEnergy. Several benchmarks show important energy reductions, e.g., AAt and CoC achieve reductions of up to 16.3\% and 11\%, respectively. These results strongly correlate with the reduction in main memory accesses, which is expected since DRAM accesses are the most energy-consuming operations on a GPU. Additionally, some of the energy savings stem from static energy reductions due to a decrease in hardware requirements, as we show later. Notice that energy efficiency is particularly crucial for mobile GPUs, as they are battery-operated, and this improvement comes at no extra cost.

\subsection{Importance of Tile Scheduler}
\label{subsec:khepri_scheduler}

\begin{figure*}[ht!]
    \centering
    \includegraphics[width=1\linewidth]{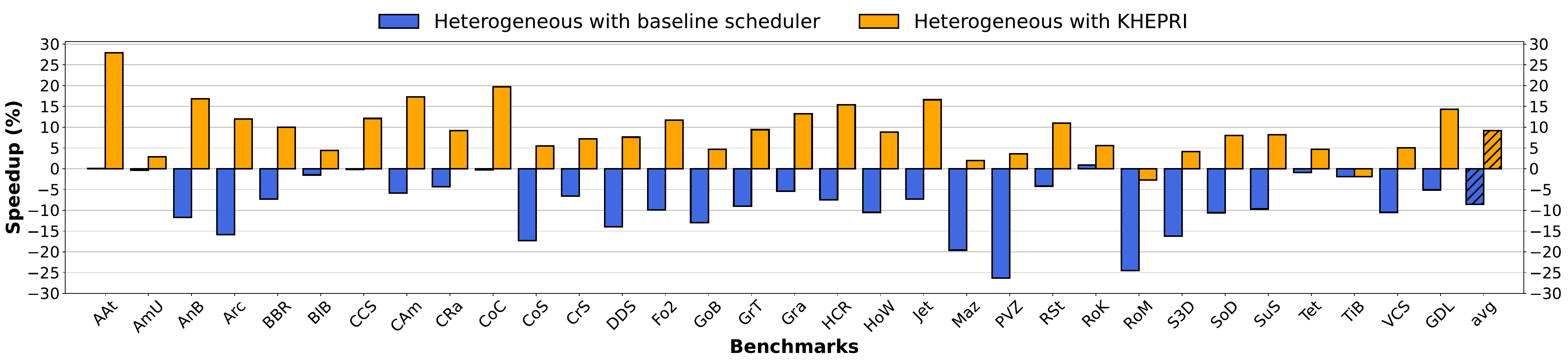}
    \caption{Speedup of a heterogeneous GPU with baseline scheduler (Z-order) and with an affinity- and locality-aware scheduler (\name) w.r.t. a traditional homogeneous GPU.}
    \label{fig:comparison_speedup_hetero_with_without_scheduler}
\end{figure*}

This subsection shows that a heterogeneous architecture performs well only when an effective scheduler is used to distribute the workload. Otherwise, the benefits may diminish, potentially leading to a negative impact on performance.

\begin{figure*}[ht!]
    \centering
    \includegraphics[width=1\linewidth]{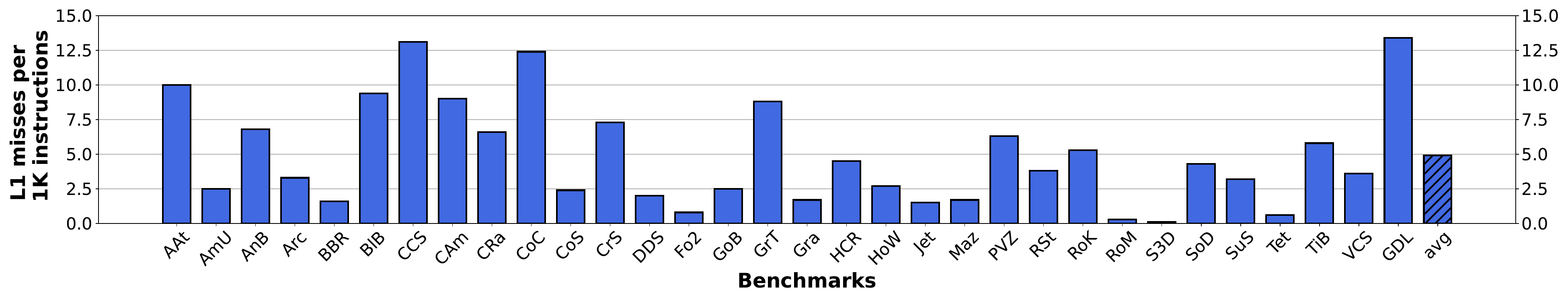}
    \caption{Number of misses per 1000 instructions in L1 texture caches for a heterogeneous GPU with a baseline scheduler (Z-order) w.r.t. \name (the lower the better).}
    \label{fig:comparison_l1_mpki_misses_hetero_with_without_scheduler}
\end{figure*}

The blue bars in \autoref{fig:comparison_speedup_hetero_with_without_scheduler} show the speedup obtained when employing the baseline (Z-order) scheduler in the heterogeneous GPU, with respect to the homogeneous GPU. In both cases, tiles are traversed following the conventional Z-order and scheduled to the least occupied Raster Unit. It can be observed that almost all applications experience a significant performance slowdown, with an average decrease of 8.6\%. In some cases, some benchmarks suffer performance degradations over 20\% (PVZ and RoM). To better put these numbers in context, the orange bars in the same figure show the speedup of \name, which we presented in the previous subsection. We can observe a considerable difference when employing an affinity- and locality-aware scheduler in a heterogeneous architecture. For example, in the case of PVZ, we observe a performance difference of nearly 30\%. On average, the difference across all benchmarks is 17.8\%. This demonstrates that heterogeneity offers significant advantages compared to a conventional homogeneous architecture, but to effectively leverage heterogeneity, it is crucial to have a tile scheduler that smartly distributes the workload (tiles).

The main reason for the poor performance of the baseline scheduler is the loss of both locality and affinity. In particular, 
\autoref{fig:comparison_l1_mpki_misses_hetero_with_without_scheduler} shows the number of L1 texture cache misses per 1000 instructions of the baseline scheduler with respect to \name, both for the same heterogeneous GPU configuration. We observe that all evaluated benchmarks show an increase in their miss rate, with an average increase of 4.9 misses per 1000 instructions. Some applications suffer a higher locality penalty (e.g., CCS and GDL), while others are less affected (e.g., S3D and RoM).
We observe that AAt shows a significant performance difference between the two schedulers of 27.8\%. This is due to an increase of 10 misses per 1000 instructions (MPKI) that is relatively small compared to its absolute MPKI (130 MPKI), which implies that workload affinity has a greater impact than locality in this benchmark. On the other hand, CrS exhibits a performance difference of 13.8\%, mainly due to an increase of 7.3 misses per 1000 instructions due to the locality loss. Summarizing, for heterogeneity to be effective, the scheduler must be aware of both the workload type and locality.

\subsection{Implementation Overheads}
\label{subsec:hw_implementation}

This subsection evaluates the timing and hardware overheads of \name. 

\noindent \textbf{Affinity computation.} To map tiles onto shader cores, the tiles are first ranked according to their memory intensity (i.e., MPKI). This sorting step uses a merge sort algorithm, which has a computational cost of $O(n\log{}n)$. The ordering logic sequentially compares pairs of values and swaps them when necessary. To support this operation, a small on-chip buffer is required to store the following per-tile data: the number of cycles spent in the Fragment Stage (16 bits), its MPKI (16 bits), and its core type affinity (1 bit). An additional 11 bits are required to store the tile ID. Conservatively assuming that two on-chip buffer reads, one comparison, and one write can be performed in three cycles, the ranking step requires an upper bound of $3 \times n\log{}n$ cycles to complete. Next, the algorithm assigns a core type to each tile based on the ranking and tile timing. Assuming that one tile is read from either the upper or the bottom part of the sorted list per cycle, and that one extra cycle is needed to accumulate its execution time to ensure a balanced time between the two types, this step requires $2 \times n$ cycles.

\noindent \textbf{Locality preservation.} To preserve locality, some tiles are reclassified, as explained in Section \ref{subsec:pres_locality}. This is achieved through a subsequent $O(n)$ pass that traverses all tiles to check their isolation status (totally or highly isolated) and reclassifies them when necessary. Afterwards, the Flood Fill algorithm is used to detect distinct regions in a frame. This algorithm traverses all the tiles and assigns each tile to a region. This process requires two queues and a boolean array that indicates whether each tile has already been visited. This procedure performs one insert and one remove operation in these two queues, one operation to mark a tile as visited, and four neighbor-check operations. Each of these operations can be performed in a single cycle, so a total of $7n$ cycles are required. The queues require 11 bits per entry.

\vspace{3mm}
Summarizing, in our experimental setup, which utilizes an FHD resolution, the entire frame is divided into 2040 tiles. Note that the number of tiles per frame depends on the screen resolution, typically totaling only a few thousand. The affinity computation and the locality preservation require a total storage cost of 16.4KB altogether, representing approximately 0.8\% of the L2 area. 

On the other hand, \name's specialized cores have a different configuration compared to the baseline (i.e., cache size, number of MSHRs, number of collector units and their crossbar, I-Buffers size, etc.). If we take this into account along with the overheads described above, we conclude that \name reduces the total GPU storage requirements by 71.6KB with respect to the baseline homogeneous architecture, which represents a reduction of slightly lower than 1\%.

Regarding timing, the affinity algorithm requires $71,365$ cycles, and the algorithm for locality preservation $18,360$ cycles, totaling $89,725$ cycles. Note, however, that each frame spends, on average, approximately $270,000$ cycles in the geometry stages for the evaluated benchmarks. Since the additional computations introduced by \name are executed in parallel with the Geometry Pipeline and require substantially less time, their latency is effectively hidden. 

\section{Related work}
\label{sec:relatedwork}

\noindent \textbf{Heterogeneous architectures.}
The diverse demands across the execution phases of modern applications have prompted researchers to investigate novel architectures that leverage this heterogeneity to enhance both performance and energy efficiency. 
This asymmetry incorporates cores of different types or complexities into a single chip to better match execution resources with each application's specific needs.
In industry, this concept has become widely adopted, with notable examples including ARM \cite{arm-dynamiq, arm-big-little, greenhalgh2011big}, Intel \cite{rotem2022intel}, and AMD \cite{amd2025hybrid, bhargava2024amd}.

Processor architectures have been extensively studied, with numerous works advocating for heterogeneous multicore designs.
Kumar et al. \cite{kumar2003single} introduced single-ISA heterogeneous architectures that combine cores with varied characteristics to maximize both performance and energy efficiency compared with their homogeneous counterparts. Composite Cores \cite{lukefahr2012composite} further improved energy efficiency by enabling both in-order and out-of-order execution within a single core. Other works in the literature have also investigated single-ISA heterogeneity \cite{kumar2004single, kumar2006core, kumar2006processor}.
Conversely, heterogeneous-ISA architectures were proposed to improve performance and energy efficiency by exploiting ISA affinity across distinct code regions within an application \cite{devuyst2012execution, venkat2014harnessing, venkat2019composite}.
Additionally, numerous previous studies have focused on runtime scheduling approaches aimed at matching workload heterogeneity with hardware heterogeneity (e.g., \cite{van2012scheduling, chen2009efficient, shelepov2009hass, van2013fairness, chen2008energy, naithani2017reliability, venkat2016hipstr, becchi2006dynamic, koufaty2010bias, lakshminarayana2009age}).

For GPUs, the exploration of heterogeneous multicore architectures remains largely unexplored.
Despite this, some GPGPU works have achieved functional specialization through techniques such as warp-specialization schemes that overlap memory access and computation \cite{cragowasp, bauer2011cudadma, bauer2014singe, wei2020efficient, wang2017decoupled, ausavarungnirun2015exploiting, kim2016warped}.
In graphics, LIBRA \cite{tomas2024libra} exploits tile diversity within frames to prevent memory-system saturation; however, this work does not employ a heterogeneous GPU architecture.
To the best of our knowledge, our work is the first to explore GPU design with heterogeneous cores for either graphics or GPGPU workloads.

\noindent \textbf{Work Scheduling.}
Regarding graphics GPU architectures, different tile traversal orders have been studied in the literature for various purposes. To achieve uniform memory utilization throughout execution, LIBRA \cite{tomas2024libra} maps tiles across different RUs by interleaving the processing of tiles with high and low memory demands. Kerbl et al. \cite{kerbl2017effective} focus on improving load balancing across threads on high-end desktop GPUs by scheduling tiles across different Graphics Processing Clusters (GPCs), each of which contains a single RU. DTexL \cite{joseph2022dtexl} employs a Hilbert tile traversal order to improve texture memory locality through quad scheduling. Another work \cite{joseph2023boustrophedonic} traverses tiles within a frame in the reverse order of the previous frame to enhance L2 texture caching. Finally, Nah et al. \cite{nah2017z2} explore mapping tiles from the left and right eyes to the same shader core for VR applications.

For general-purpose GPU architectures (GPGPUs), some previous studies focus on warp scheduling strategies aimed at improving cache locality \cite{rogers2012cache, zhang2018locality, lee2015cawa, li2015priority, rogers2013divergence, lee2014caws, wang2016oaws, sethia2015mascar, jog2013owl}, hiding long-latency memory accesses \cite{narasiman2011improving, jog2013orchestrated, oh2016apres, dublish2019poise}, alleviating branch divergence \cite{fung2011thread, meng2010dynamic}, and reducing synchronization barriers among warps \cite{liu2015saws}. Some works also propose register-bank-aware warp schedulers to minimize bank conflicts \cite{barnes2023mitigating} or enhance register file energy efficiency \cite{gebhart2011energy}. In addition, several proposals explore thread-block (CTA) scheduling techniques to improve cache reuse \cite{li2017locality, lee2014improving, chen2017improving, tripathy2021paver, wang2016laperm, huzaifa2020inter}, mitigate long memory latencies \cite{kayiran2013neither}, or distribute CTAs across chiplets \cite{arunkumar2017mcm} or multiple GPUs \cite{khairy2020locality, kim2018coda} to improve locality and reduce remote memory accesses.

\noindent \textbf{Locality.}
One of the main goals of \name is the preservation of locality. Since texture accesses are a primary source of traffic within the memory hierarchy and typically consume the majority of DRAM bandwidth, many works have focused on improving texture locality. DTM-NUCA \cite{corbalan2022dtm} proposes a NUCA organization to enhance the effective capacity of L1 texture caches. Ukarande et al. \cite{ukarande2021locality} map screen-adjacent CTAs to the same SM to maximize the locality of texture access patterns. Xie et al. \cite{xie2017processing} explore the use of PIM architectures to reduce DRAM traffic caused by texture accesses. Other works focus on prefetching textures into L1 texture caches \cite{arnau2012boosting, igehy1998prefetching} or applying texture compression \cite{nystad2012adaptive, strom2005packman, xiao2012self, fenney2003texture, akenine2003graphics}. Notably, all of these works are orthogonal to our proposal.

\section{Conclusion}
\label{sec:conclusions}

While heterogeneous architectures have been extensively studied for general-purpose processors, they remain largely unexplored in the context of GPUs. Motivated by the opportunities presented by different execution phases in graphical applications, this work introduces \name, a heterogeneous GPU architecture specifically designed for rendering graphics applications. Our proposal employs two types of GPU shader cores: one core type specialized for sustaining higher levels of ILP, and another core type optimized to handle more memory requests simultaneously (MLP). To map tiles onto cores, \name incorporates a novel affinity- and locality-aware tile scheduler, which is a key component to make an effective use of heterogeneity. The tile scheduler predicts the memory intensity of each tile within a frame by leveraging frame-to-frame coherence to determine the most suitable core type. Since locality is a key aspect, the scheduler includes a locality-aware mechanism that ensures that tile assignment does not negatively impact the performance of the memory hierarchy. Furthermore, to accommodate the varying behavior and runtime of different applications, the tile scheduler dynamically adapts its mapping decisions on a per-frame basis using collected tile statistics. 

Our experimental evaluation shows that GPU core heterogeneity benefits a wide range of real-world graphics applications, yielding significant improvements in both performance (\resultPerformance) and energy (\resultEnergy), without incurring any overhead.


\section*{Acknowledgment}
This work has been supported by the CoCoUnit ERC Advanced Grant of the EU’s Horizon 2020 program (grant No 833057), the Spanish State Research Agency (MCIU/AEI/10.13039/501100011033) and FEDER/UE under grants PID2020-113172RB-I00 and PID2024-155476OB-I00, the Catalan Agency for University and Research (AGAUR) under grant 2021SGR00383, the ICREA Academia program and the FPI research grant PRE2021-100336.

\nocite{bstctl:nodash}
\bibliographystyle{IEEEtranS}
\bibliography{refs}

\end{document}